\documentclass[12pt,a4paper]{article}
\usepackage{amsmath}
\usepackage{amssymb}
\usepackage{amscd}

\def\h{\hbar}

\begin{document}

\begin{titlepage}
\title{
\hfill\parbox{4cm}
{\normalsize KUNS-1712\\{\tt hep-th/0103018}}\\
\vspace{1cm}
Fuzzy Sphere and Hyperbolic Space\\ from Deformation Quantization}
\author{
Isao {\sc Kishimoto}\thanks{{\tt
    ikishimo@gauge.scphys.kyoto-u.ac.jp}}
\\[7pt]
{\it Department of Physics, Kyoto University, Kyoto 606-8502, Japan}
}
\date{\normalsize March, 2001}
\maketitle
\thispagestyle{empty}

\begin{abstract}
\normalsize
We explicitly construct noncommutative $*$ products on circularly
symmetric two dimensional space by using the technique of  Fedosov's
deformation quantization. Especially, on constant curvature spaces
i.e., $S^2$ and $H^2$, we get $su(2)$ and $su(1,1)$ algebra
respectively. These are candidates of $*$ products applicable to
noncommutative field theories or noncommutative gauge theories on
spaces with nontrivial symplectic structure.
\end{abstract}

\end{titlepage}

\section{Introduction}

Since the relation between string theory and noncommutative geometry
was discussed in \cite{SW}, noncommutative field theories and
noncommutative gauge theories have been investigated enthusiastically
from various viewpoints.

Many authors use the Moyal product\footnote{Here we call
        $*=\exp\left({i\over 2}{{\overleftarrow\partial}\over\partial
        x^i}\theta^{ij}{{\overrightarrow \partial}\over\partial
        x^j}\right)$ with constant $\theta^{ij}=-\theta^{ji}$ the
        Moyal product.}
as noncommutative associative  $*$ product for explicit calculations. 
It corresponds to a constant NS-NS $B$-field background in flat space in
        the context of string theory. 
On the other hand, at least formally, more general $*$ products
which may correspond to string theory on nonconstant $B$-field
        background in curved space 
are defined by some authors\footnote{\cite{KON},\cite{Fedbk}, for
        example.}.
However, explicit form of $*$ products other than the Moyal product has
        been scarcely discussed in physical context\footnote{
In \cite{CS}, nonassociative star product which generalizes
        \cite{KON},\cite{Fedbk} is discussed to describe D-brane in
        curved backgrounds.}.

In this paper, we use the technique of Fedosov's deformation
quantization \cite{Fedbk} to get explicit forms of $*$ products on
nontrivial backgrounds. 
For simplicity, we investigate $*$ products on circularly symmetric two
dimensional spaces. Specifically, we focus on constant curvature
spaces $S^2,H^2$ and ${\mathbb R}^2$, and explicitly construct $*$
products which are different from the Moyal product.
We also discuss some physical applications of our $*$ products.

\section{Construction of $*$ product\label{sec:CIR}}

Here we review the construction of Fedosov's $*$ product very briefly\footnote{
See \cite{Fedbk},\cite{AK2} for details.
}, and apply this procedure to circularly symmetric two dimensional spaces.

First, for a given symplectic manifold $(M,\Omega_0)$, we define the Weyl
algebra bundle $W$ which has $\circ$ product of the Moyal type and its
Abelian connection $D$ with some input parameter.
For ${\rm Ker}D\subset W$ (which is called flat section $W_D$), we get
a one to one correspondence with $C^\infty(M)[[\hbar]]$, where $\hbar$ is the deformation parameter. 
We denote the map from $C^\infty(M)[[\hbar]]$  to $W_D$ as $Q$, and its inverse map as $\sigma$.
Then Fedosov's $*$ product on $C^\infty(M)[[\hbar]]$ is defined by
\begin{equation}
\label{eqn:STARDF}
a_0*b_0:=\sigma(Q(a_0)\circ Q(b_0)),\ \ a_0,b_0\in C^\infty(M)[[\hbar]].
\end{equation}
This is a solution of the problem of deformation quantization, i.e.,\\
$*$ is associative and its commutator $[\ ,\ ]_*$ is expanded as
\begin{equation}
[\ ,\ ]_*=i\hbar \{\ ,\ \}+{\cal O}(\hbar^2)
\end{equation}
where $\{\ ,\ \}$ is the Poisson bracket with respect to the symplectic
form $\Omega_0$.

Now, we apply this procedure to a two dimensional space $M$ with metric
\begin{equation}
\label{eqn:MET}
ds^2=e^{\Phi(r)}(dr^2+r^2d\theta^2),
\end{equation}
where $\Phi(r)$ is some function of $r$ only (i.e. circularly
symmetric space) for simplicity.
Its volume form is given by
\begin{equation}
\Omega_0=e^{\Phi(r)}rdr\wedge d\theta,
\end{equation}
and we identify it with symplectic form. Using Fedosov's procedure with the input\footnote{
See \cite{Fedbk},\cite{AK2} for the meaning of $\nabla,\Omega_1,\mu,\delta$.
Here we choose these parameters in such a way that the iteration formula
(eq.(21) of \cite{AK2}) which gives an Abelian connection is satisfied
trivially, i.e., $\nabla{\rm r}+{i\over \h}{\rm r}\circ{\rm r}=0$. Then 
we get ${\rm
  r}=\delta\mu+\delta^{-1}(d(\omega_{ij}y^i\theta^j)-\Omega_1)$ for
the input (\ref{eqn:IN}).
}
\begin{eqnarray}
\label{eqn:IN}
&&\Omega_0=\theta^1\wedge\theta^2=-{1\over2}\omega_{ij}\theta^i\wedge\theta^j,\nonumber\\
&&\theta^1=e^{\Phi(r)}dr,\quad \theta^2=rd\theta,\quad
\omega_{ij}=\left(
\begin{array}{cc}
0  &  -1  \\
1  &   0
\end{array}
\right),\nonumber \\
&&\Omega_1=0,\ \nabla=d, \nonumber \\
&&\mu={1\over3}e^{-\Phi(r)}r^{-1}(y^1)^2y^2,
\end{eqnarray}
we get an Abelian connection $D$ as 
\begin{eqnarray}
&&Da=da-\delta a+{i\over\hbar}({\rm r}\circ a-a\circ {\rm r}),\quad a\in W,\nonumber \\
&&{\rm r}=e^{-\Phi(r)}r^{-1}y^1y^2\theta^1,\nonumber \\
&&\circ :=\exp\left(-{i\hbar\over2}{{\overleftarrow \partial}\over\partial y^i}\omega^{ij}{{\overrightarrow \partial}\over\partial y^j}\right),\quad \omega^{ij}:=(\omega^{-1})^{ij}.
\end{eqnarray}
For this Abelian connection $D$, we solve the equation $Da=0$ and get the map $Q:C^\infty(M)[[\hbar]]\rightarrow W_D$ as
\begin{equation}
\label{eqn:FLAT}
a=Q(a_0(r,\theta))=a_0\left(G(r,y^1),\theta+{y^2\over r}\right),
\end{equation}
where $G(r,y^1)$ is given by
\begin{equation}
\int_r^{G(r,y^1)}e^{\Phi(r')}r'dr'=y^1 r.
\end{equation}
Then we can define a $*$ product on $M$ by eq.(\ref{eqn:STARDF}).

\section{$S^2$ case\label{sec:S2}}
In this section we apply the result of \S\ref{sec:CIR} to the case $M=S^2$.
We consider 2-sphere $S^2$ with radius $R$, which is defined as two dimensional surface embedded in ${\mathbb R}^3$:
\begin{equation}
\label{eqn:S2DF}
(X^1)^2+(X^2)^2+(X^3)^2=R^2.
\end{equation}
We parametrize the coordinate $X^i,i=1,2,3$ on $S^2$ as 
\begin{eqnarray}
\label{eqn:X123DF}
&&X^1={2R^2r\over r^2+R^2}\cos\theta,\ X^2={2R^2r\over
  r^2+R^2}\sin\theta,\ X^3=R{r^2-R^2\over r^2+R^2},\nonumber\\
&&r\geq0,\ 0\leq\theta\leq2\pi.
\end{eqnarray}
Then the metric of $S^2$, $ds^2=(dX^1)^2+(dX^2)^2+(dX^3)^2$, is given by
\begin{equation}
\label{eqn:S2DS2}
ds^2={4R^4\over(r^2+R^2)^2}(dr^2+r^2d\theta^2),
\end{equation}
and the conformal factor $e^{\Phi}$ of eq.(\ref{eqn:MET}) is
identified as
\begin{equation}
\label{eqn:FS2}
e^{\Phi(r)}={4R^4\over(r^2+R^2)^2}.
\end{equation}
From eqs.\ (\ref{eqn:FS2}), (\ref{eqn:FLAT}) and (\ref{eqn:STARDF}),
we get the  explicit form of our  $*$ product on $S^2$:
\begin{eqnarray}
\label{eqn:AS2STR}
&&a_0(r,\theta)*b_0(r,\theta)\nonumber\\
&=&\biggl(a_0\left(\sqrt{r^2+{y^1\over2R^2}r(r^2+R^2)\over1-{y^1\over2R^4}r(r^2+R^2)},\theta+{y^2\over
    r}\right)\exp\left(-{i\h\over2}\left({{\overleftarrow
        \partial}\over\partial y^1}{{\overrightarrow
        \partial}\over\partial y^2}-{{\overleftarrow
        \partial}\over\partial y^2}{{\overrightarrow
        \partial}\over\partial y^1}\right)\right)\nonumber\\
&&\cdot
b_0\left(\sqrt{r^2+{y^1\over2R^2}r(r^2+R^2)\over1-{y^1\over2R^4}r(r^2+R^2)},\theta+{y^2\over
    r}\right)\biggr)_{y^1=y^2=0}.
\end{eqnarray}
By using this definition, we can calculate $*$ product of the $S^2$
coordinate $X^i$ (\ref{eqn:X123DF}). In particular, we have
\begin{eqnarray}
\label{eqn:FUZZYS2}
&&[X^i,X^j]_*=i{\h\over R}\varepsilon^{ijk}X^k,\\
\label{eqn:FZS2}
&&X^1*X^1+X^2*X^2+X^3*X^3=R^2\left(1-{\h^2\over4R^4}\right),
\end{eqnarray}
where $\varepsilon^{ijk}$ is the antisymmetric tensor with
$\varepsilon^{123}=+1$.
Eq.(\ref{eqn:FUZZYS2}) means that the commutators of $X^i$'s form
$su(2)$ algebra which is known as fuzzy sphere algebra, and
eq.(\ref{eqn:FZS2}) means that its radius is given by
$R\sqrt{1-{\h^2\over4R^4}}$
which is deformed by ${\cal O}(\h^2)$ from the original radius $R$ of
commutative $S^2$ (\ref{eqn:S2DF}).
Namely, we have obtained a fuzzy sphere by deforming $S^2$ using the $*$ product (\ref{eqn:AS2STR}).

\section{$H^2$ case\label{sec:H2}}

In this section we apply the result of \S\ref{sec:CIR}
to the case $M=H^2$.
Calculation is quite similar to the $S^2$ case (\S\ref{sec:S2}).
We consider two dimensional hyperbolic space $H^2$ with radius $R$,
which is defined as two dimensional surface embedded in ${\mathbb
  R}^{1,2}$:
\begin{equation}
\label{eqn:H2DF}
-(Y^0)^2+(Y^1)^2+(Y^2)^2=-R^2,\quad Y^0>0.
\end{equation}
We parametrize the coordinates $Y^i,i=0,1,2$ on $H^2$ as
\begin{eqnarray}
\label{eqn:Y123DF}
&&Y^0=R{R^2+r^2\over R^2-r^2},\ Y^1={2R^2r\over R^2-r^2}\cos\theta,\ Y^2={2R^2r\over R^2-r^2}\sin\theta,\ \nonumber\\
&&0\leq r\leq R,\ 0\leq\theta\leq2\pi.
\end{eqnarray}
Then, the metric of $H^2$, $ds^2=-(dY^0)^2+(dY^1)^2+(dY^2)^2$, and the
conformal factor are given respectively by
\begin{eqnarray}
\label{eqn:H2DS2}
&&ds^2={4R^4\over(R^2-r^2)^2}(dr^2+r^2d\theta^2),\\
\label{eqn:FH2}
&&e^{\Phi(r)}={4R^4\over(R^2-r^2)^2}.
\end{eqnarray}
From eqs.\ (\ref{eqn:FH2}), (\ref{eqn:FLAT}) and (\ref{eqn:STARDF}),
we get the explicit form of our $*$ product on $H^2$:
\begin{eqnarray}
\label{eqn:AH2STR}
&&a_0(r,\theta)*b_0(r,\theta)\nonumber\\
&=&\biggl(a_0\left(\sqrt{r^2+{y^1\over2R^2}r(R^2-r^2)\over1+{y^1\over2R^4}r(R^2-r^2)},\theta+{y^2\over
    r}\right)\exp\left(-{i\h\over2}\left({{\overleftarrow
        \partial}\over\partial y^1}{{\overrightarrow
        \partial}\over\partial y^2}-{{\overleftarrow
        \partial}\over\partial y^2}{{\overrightarrow
        \partial}\over\partial y^1}\right)\right)\nonumber\\
&&\cdot
b_0\left(\sqrt{r^2+{y^1\over2R^2}r(R^2-r^2)\over1+{y^1\over2R^4}r(R^2-r^2)},\theta+{y^2\over
    r}\right)\biggr)_{y^1=y^2=0}.
\end{eqnarray}
By using this definition, we obtain the following  $*$ products of the $H^2$
coordinate $Y^i$ (\ref{eqn:Y123DF}):
\begin{eqnarray}
\label{eqn:FUZZYH2}
&&[Y^0,Y^1]_*=i{\h\over R}Y^2,\quad [Y^2,Y^0]_*=i{\h\over R}Y^1,\quad
[Y^1,Y^2]_*=-i{\h\over R}Y^0,\\
\label{eqn:FZH2}
&&-Y^0*Y^0+Y^1*Y^1+Y^2*Y^2=-R^2\left(1-{\h^2\over4R^4}\right).
\end{eqnarray}
Eq.(\ref{eqn:FUZZYH2}) means that commutators of $Y^i$'s form $su(1,1)$
algebra which corresponds to isometry of $H^2$, and eq.(\ref{eqn:FZH2})
means that its radius is given by $R\sqrt{1-{\h^2\over4R^4}}$ which is
deformed by ${\cal O}(\h^2)$ from the original radius $R$ of commutative
$H^2$ (\ref{eqn:H2DF}).
Namely, we get fuzzy hyperbolic space by deforming  $H^2$ using the $*$
product (\ref{eqn:AH2STR}).

\section{Large $R$ limit and ${\mathbb R}^2$}

Here we consider large radius limit of the results of \S\ref{sec:S2}
and \S\ref{sec:H2}.
The sectional curvature of $S^2$ (\ref{eqn:S2DF})
($H^2$ (\ref{eqn:H2DF})) is ${1\over R^2}$ ($-{1\over R^2}$),
which tends to $+0$ ($-0$) in the limit $R\rightarrow\infty$.
Therefore they approach the flat space ${\mathbb R}^2$ in the large
$R$ limit in the usual commutative picture.
How about it from the noncommutative viewpoint?

For comparison, we construct a $*$ product on ${\mathbb R}^2$
following the method of \S\ref{sec:CIR}.
We adopt as its flat metric
\begin{equation}
\label{eqn:R2METRIC}
ds^2=4(dr^2+r^2d\theta^2)
\end{equation}
with its front factor 4 chosen so that (\ref{eqn:R2METRIC}) coincides with
the large $R$ limit of (\ref{eqn:S2DS2}) and (\ref{eqn:H2DS2}).
With $e^\Phi=4$, we get the explicit form of our $*$ product on
${\mathbb R}^2$:
\begin{eqnarray}
\label{eqn:AR2STR}
&&a_0(r,\theta)*b_0(r,\theta)\nonumber\\
&&=\biggl(a_0\left(\sqrt{r^2+{y^1r\over2}},\theta+{y^2\over
    r}\right)\exp\left(-{i\h\over2}\left({{\overleftarrow
        \partial}\over\partial y^1}{{\overrightarrow
        \partial}\over\partial y^2}-{{\overleftarrow
        \partial}\over\partial y^2}{{\overrightarrow
        \partial}\over\partial y^1}\right)\right)\nonumber\\
&&\cdot b_0\left(\sqrt{r^2+{y^1r\over2}},\theta+{y^2\over
    r}\right)\biggr)_{y^1=y^2=0}.
\end{eqnarray}
Then, we can calculate the  $*$ products of the complex coordinate
$z:=re^{i\theta},\ {\bar z}:=re^{-i\theta}$:
\begin{eqnarray}
\label{eqn:AR2Z}
&&z*z=\sqrt{r^4-{\h^2\over16}}e^{2i\theta}=\overline{{\bar z}*{\bar z}},\quad
z*{\bar z}=r^2-{\h\over4},\quad {\bar z}*z=r^2+{\h\over4},\nonumber\\
&&[z,{\bar z}]_*=-{\h\over2}.
\end{eqnarray}
The commutator $[z,{\bar z}]_*$ coincides
with that of the usual Moyal product for Cartesian coordinates on
${\mathbb R}^2$, but $*$ product itself is different from the Moyal
product. This difference comes from ambiguity of deformation
quantization.

We can calculate the  commutator $[z,{\bar z}]_*$ also in the $S^2$ and
$H^2$ cases. 
For $S^2$, from eq.(\ref{eqn:AS2STR}) we get
\begin{equation}
\label{eqn:S2ZB}
[z,{\bar
  z}]_*={-{\h\over2R^4}(r^2+R^2)^2\over1-\left({\h\over4R^4}(r^2+R^2)\right)^2}=-{\h\over2R^4}(R^2+z*{\bar z})(R^2+{\bar z}*z).
\end{equation}
And for $H^2$, from eq.(\ref{eqn:AH2STR}) we get
\begin{equation}
\label{eqn:H2ZB}
[z,{\bar
  z}]_*={-{\h\over2R^4}(R^2-r^2)^2\over1-\left({\h\over4R^4}(R^2-r^2)\right)^2}=-{\h\over2R^4}(R^2-z*{\bar z})(R^2-{\bar z}*z).
\end{equation}
Both eqs.(\ref{eqn:S2ZB}) and (\ref{eqn:H2ZB}) are reduced to
  $[z,{\bar   z}]_*=-{\h\over2}$ (\ref{eqn:AR2Z}) as $R\to\infty$.
In other words, the $*$ product which we obtained in \S\ref{sec:CIR}
connects $su(2)$ algebra (or fuzzy $S^2$) with $su(1,1)$ algebra (or
fuzzy $H^2$) through $R=\infty$.

\section{An application
\label{sec:APP}}

In the previous sections, we explicitly calculated $*$ products by
using Fedosov's formulation. They are candidates of $*$ product for
defining noncommutative field theory or noncommutative gauge theory on
fuzzy $S^2,H^2$ and ${\mathbb R}^2$.

As an example, we discuss four dimensional noncommutative $U(1)$ gauge
theory with one scalar field which is given by the action\footnote{
The symbol ${\rm Tr}$ is trace for the $*$ product satisfying ${\rm
  Tr}f*g={\rm Tr}g*f$ \cite{Fedbk}, but we can discuss equations of
motion without using the explicit form of the trace.
}
\begin{equation}
\label{eqn:ACTION}
S={\rm
  Tr}\left({1\over4}G^{IJ}G^{KL}F_{IK}*F_{JL}+{1\over2}G^{IJ}D_I\phi*D_J\phi\right).
\end{equation}
We assume that only two dimensional space is noncommutative (1,2
direction), 
and use a general formulation of noncommutative gauge theory of \cite{AK2}:
\begin{eqnarray}
&&G^{IJ}=\delta^{IJ},\ I,J=1,\cdots,4,\nonumber\\
&&F_{IJ}=\partial_IA_J-\partial_JA_I-i[A_I,A_J]_*-{J_{IJ}\over\h},\quad
J_{12}=-J_{21}=1,{\rm others}=0,\nonumber\\
&&\partial_I={i\over\h}[-J_{IJ}{\tilde \phi}^J,\ ]_*,\ I=1,2,\ \ \partial_3={\partial\over\partial x^3},\partial_4={\partial\over\partial x^4}\nonumber\\
&&D_I\phi=\partial_I\phi-i[A_I,\phi]_*,
\end{eqnarray}
Here,  ${\tilde \phi}^I$ is the ``canonical'' noncommutative coordinate
satisfying
\begin{equation}
\label{eqn:CAN}
{i\over\h}[{\tilde \phi}^1,{\tilde \phi}^2]_*=1.
\end{equation}
Its explicit form is 
\begin{equation}
{\tilde \phi}^1={2Rr\over\sqrt{r^2+R^2}}\cos \theta,\ {\tilde \phi}^2={2Rr\over\sqrt{r^2+R^2}}\sin \theta 
\end{equation}
for fuzzy $S^2$ (\ref{eqn:AS2STR}),
\begin{equation}
{\tilde \phi}^1={2Rr\over\sqrt{R^2-r^2}}\cos \theta,\ {\tilde \phi}^2={2Rr\over\sqrt{R^2-r^2}}\sin \theta 
\end{equation}
for fuzzy $H^2$ (\ref{eqn:AH2STR}), and
\begin{equation}
\label{eqn:PHIR2}
{\tilde \phi}^1=2r\cos \theta,\ {\tilde \phi}^2=2r\sin \theta 
\end{equation}
for fuzzy ${\mathbb R}^2$ (\ref{eqn:AR2STR}).
The action (\ref{eqn:ACTION}) is invariant under noncommutative $U(1)$ 
gauge transformation:
\begin{equation}
\delta_\lambda A_I=\partial_I\lambda-i[A_I,\lambda]_*,\quad\quad
\delta_\lambda\phi=-i[\phi,\lambda]_*.
\end{equation}

The equations of motion of (\ref{eqn:ACTION}) are
\begin{equation}
D^IF_{IJ}=-i[\phi,D_J\phi]_*,\ D^ID_I\phi=0,
\end{equation}
and we obtain a solution by solving the $U(1)$ noncommutative BPS
equation:
\begin{equation}
\label{eqn:BPS}
B_I= D_I\phi,I=1,2,3,\ \ \partial_4=0,A_4=0,\ \
B_I:={1\over2}\varepsilon^{IJK}\left(F_{JK}+{J_{JK}\over\h}\right).
\end{equation}
Under the ansatz
\begin{eqnarray}
\label{eqn:ANSATZ}
&&A_1+iA_2=if_A(l,x^3)({\tilde \phi}^1+i{\tilde \phi}^2),\quad\quad
A_3=0,\nonumber\\
&&\phi=f(l,x^3),\quad\quad l:=\sqrt{({\tilde \phi}^1)^2+({\tilde
    \phi}^2)^2+(x^3)^2},
\end{eqnarray}
eq.(\ref{eqn:BPS}) can be rewritten as
\begin{eqnarray}
\label{eqn:ITE}
&&\partial_3G^{(m)}-4\partial_Lf^{(m)}
=\sum_{
\begin{subarray}{c}
2n+k=m,\\
n\geq1
\end{subarray}
}{4\partial_L^{2n+1}f^{(k)}\over(2n+1)!}+\sum_{
\begin{subarray}{c}
2n+k+k'\\
=m-1
\end{subarray}
}{4G^{(k')}\partial_L^{2n+1}f^{(k)}\over(2n+1)!},\nonumber\\
&&\partial_3f^{(m)}-\partial_L(LG^{(m)})=\sum_{
\begin{subarray}{c}
2n+k=m,\\
n\geq1
\end{subarray}
}{\partial_L^{2n+1}(LG^{(k)})\over(2n+1)!}
\end{eqnarray}
with
\begin{equation}
L:=({\tilde \phi}^1)^2+({\tilde \phi}^2)^2,\quad
f=\sum_{k=0}^\infty\h^k f^{(k)},\quad
\left({1\over\h}+f_A\right)^2={1\over\h^2}+{1\over\h}\sum_{k=0}^\infty\h^k G^{(k)}.
\end{equation}
We can solve eq.(\ref{eqn:ITE}) order by order in $\h$, and we get
\begin{eqnarray}
&&f={g\over l}+\h g^2\left({2x^3\over l^4}-{1\over
    l^3}\right)+\h^2\left({-8g^3x^3\over
    l^6}-{g\over4l^5}+\left({5g\over8}+10g^3\right){(x^3)^2\over
    l^7}\right)+{\cal O}(\h^3),\nonumber\\
&&f_A={g\over l(l+x^3)}+\h g^2\left({2\over l^4}-{1\over
    l^3(l+x^3)}-{1\over2l^2(l+x^3)^2}\right)\nonumber\\
&&+\h^2\biggl({-8g^3\over l^6}+{4g^3\over l^5(l+x^3)}+{g^3\over
    l^4(l+x^3)^2}+{g^3\over2l^3(l+x^3)^3}-\left({5g\over8}+10g^3\right){x^3\over l^7}\biggr)+{\cal O}(\h^3),\nonumber\\
\end{eqnarray}
as a solution such that it becomes the $U(1)$ Dirac monopole in
the commutative limit (i.e., $\h\rightarrow0$). In the fuzzy
 ${\mathbb R}^2$ 
  case (\ref{eqn:PHIR2}), the ${\cal O}(\h)$ terms coincide with those in
  \cite{HH} which solved the equations of motion with the usual Moyal product.

\section{Conclusion and discussion}

In this paper we have presented explicit construction of $*$
products on two dimensional constant curvature spaces $S^2,H^2$ and ${\mathbb
  R}^2$.
We have found that the algebras of the $*$ products represent fuzzy 
$S^2,H^2$ and ${\mathbb R}^2$ because the commutators of the $*$ product form
$su(2),su(1,1)$ and Heisenberg algebra respectively.  
The commutators $[z,{\bar z}]_*$ for fuzzy 
$S^2$ and $H^2$ are reduced to that of fuzzy ${\mathbb R}^2$ in the large $R$
limit.
In this sense, fuzzy $S^2$ and $H^2$ approach to fuzzy ${\mathbb
  R}^2$ as $R\to\infty$.
This is consistent with usual commutative picture.

In \S \ref{sec:APP} we applied explicit form of our $*$ products to $U(1)$
noncommutative BPS equation (\ref{eqn:BPS}), and obtained its solution
to ${\cal O}(\h^2)$. 
In eq.(\ref{eqn:BPS}) the $*$ product appears only in the commutator $[\ ,\
]_*$. Therefore, eq.(\ref{eqn:BPS}) is solved unifiedly for
fuzzy $S^2,H^2$ and ${\mathbb R}^2$ by using ``canonical'' noncommutative
coordinate ${\tilde \phi}^I$ (\ref{eqn:CAN}). In other words, we can
get a solution of eq.(\ref{eqn:BPS}) even if the definition of $*$ is
different as long as we use ``canonical'' noncommutative coordinate
${\tilde \phi}^I$ for the $*$ product.

To study the effects of the difference of $*$ products themselves, we
should consider noncommutative equations containing ``bare'' $*$
products.
Its typical example is $\phi*\phi=\phi$ which is essentially the equation
for noncommutative soliton \cite{GMS}. Even for the ${\mathbb R}^2$ case,
the $*$ product which we get here is different from the usual Moyal product,
and hence $\phi\sim\exp(- r^2)$ is {\it not} a solution\footnote{
In the case of the Moyal product, this is a solution.
} of  $\phi*\phi=\phi$.
It is a future problem to find an explicit solution of it and to
investigate its meaning.

For fuzzy $S^2$, $*$ product is usually defined by using
representation matrix of $su(2)$ and spherical harmonic function,
and depends on the size of matrix. On the other hand our $*$ product
depends on the deformation parameter $\h$, so they
are very different in appearance.
It is also a future problem to study an explicit relation between them.
If the relation becomes clear, our $*$ product may give some
suggestions to string theory in the literature \cite{SHOME} for example.

\section*{Acknowledgements}
We would like to thank T.~Asakawa, S.~Goto, H.~Hata, H.~Kawai,
S.~Moriyama and S.~Terashima for valuable discussions and comments.
This work is supported in part by the Grant-in-Aid for Scientific
Research (\#9858) from the Ministry of Education, Science, Sports and
Culture of Japan.


\begin{thebibliography}{99}
\bibitem{SW}
N.~Seiberg and E.~Witten,
{\it ``String Theory and Noncommutative Geometry,''}
{\sl JHEP}~{\bf 9909} (1999) 032.
{\tt hep-th/9908142}.
\bibitem{KON}
M.~Kontsevich,
{\it ``Deformation quantization of Poisson manifolds, I,''}
{\tt q-alg/9709040}.
\bibitem{Fedbk}
B.V.~Fedosov,
{\it ``Deformation quantization and index theory,''}
{\sl  Berlin, Germany: Akademie-Verl.} (1996).
\bibitem{CS}
L.~Cornalba and R.~Schiappa,
{\it ``Nonassociative Star Product Deformations for D-brane
  Worldvolumes in Curved Backgrounds,''}
{\tt hep-th/0101219}.
\bibitem{AK2}
T.~Asakawa and I.~Kishimoto,
{\it ``Noncommutative Gauge Theories from Deformation Quantization,''}
{\sl Nucl.~Phys.}~{\bf B591} (2000) 611-635. {\tt hep-th/0002138}.
\bibitem{HH}
K.~Hashimoto and T.~Hirayama,
{\it ``Branes and BPS Configurations of Non-Commutative/Commutative Gauge
Theories,''}
{\sl Nucl.~Phys.}~{\bf B587} (2000) 207-227.
{\tt hep-th/0002090}.
\bibitem{GMS}
R.~Gopakumar, S.~Minwalla and A.~Strominger,
{\it ``Noncommutative Solitons,''}
{\sl JHEP}~{\bf 0005} (2000) 020.
{\tt hep-th/0003160}.
\bibitem{SHOME}
A.~Yu.~Alekseev, A.~Recknagel and  V.~Schomerus,
{\it ``Brane Dynamics in Background Fluxes and Non-commutative
Geometry,''}
{\sl JHEP}~{\bf 0005} (2000) 010.
{\tt hep-th/0003187}.
\end{thebibliography}
\end{document}